  \documentclass[12pt,onecolumn]{IEEEtran}

\usepackage{url}
\usepackage{amsmath}
\usepackage{amsfonts}
\usepackage{theorem}
\usepackage[dvips]{epsfig}      
\usepackage{enumerate}
\usepackage{color}                  

\usepackage{verbatim}

\usepackage{amssymb}

\usepackage{float}

\theoremstyle{plain}
\newtheorem{Theorem}{Theorem}

\newtheorem{Proposition}{Proposition}
\newtheorem{Corollary}{Corollary}

\newtheorem{Problem}{Problem}

{\theorembodyfont{\rmfamily} \newtheorem{Example}{Example}}
\newtheorem{Remark}{Remark}

\newcommand {\R}{\mathbb R}

\newcommand{\be}{\begin{equation}}
\newcommand{\ee}{\end{equation}}
\newcommand{\Int}{\operatorname{{\mathrm int}}}

{\theorembodyfont{\rmfamily} \newtheorem{Definition}{Definition}}

\newcommand{\LMD}{\lambda_0,\dots,\lambda_n}

\newcommand{\BLMD}{\bar \lambda_0,\dots,\bar \lambda_n}

\begin{document}

 \title{Optimal Down Regulation of mRNA Translation\thanks{The research of MM and TT is partially supported by a research grant from  the Israeli Ministry of Science, Technology, and Space.
The research of MM is also supported by a research grant from the Israel Science Foundation}}
\author{ Yoram Zarai,   Michael Margaliot, and Tamir Tuller \IEEEcompsocitemizethanks{
\IEEEcompsocthanksitem
 Y. Zarai is with the School of Electrical Engineering, Tel-Aviv
University, Tel-Aviv 69978, Israel.
E-mail: yoramzar@mail.tau.ac.il
\IEEEcompsocthanksitem
M. Margaliot is with the School of Electrical Engineering and the Sagol School of Neuroscience, Tel-Aviv
University, Tel-Aviv 69978, Israel.
E-mail: michaelm@eng.tau.ac.il
\IEEEcompsocthanksitem
  T. Tuller is with the Dept. of Biomedical Engineering and the Sagol School of Neuroscience, Tel-Aviv
University, Tel-Aviv 69978, Israel.
E-mail: tamirtul@post.tau.ac.il
 }}


\maketitle

 \begin{abstract}

Down regulation of mRNA translation is an important problem in various bio-medical domains ranging from developing effective medicines for tumors and for viral diseases to developing attenuated virus strains that can be used for vaccination.
Here, we study the problem of down regulation of mRNA translation using a mathematical model called the ribosome flow model~(RFM). In the RFM, the mRNA molecule is modeled as a chain of $n$ sites. The flow of ribosomes between consecutive sites is regulated by $n+1$ transition rates.
	Given a set of feasible transition rates, that models the outcome of all possible mutations, we consider the problem of maximally  down regulating the translation rate by altering the  rates within this set of feasible rates.  Under certain conditions on the feasible set, we show that an optimal solution can be determined efficiently.
We also rigorously analyze two special cases of the down regulation optimization problem. Our results suggest that one must focus on the position along the mRNA molecule where the transition rate has the strongest effect on the protein production rate. However, this rate is not necessarily the slowest transition rate along the mRNA molecule. We discuss some of the biological implications of these results.

  \end{abstract}

\section*{Introduction}
Gene expression is the process by which the genetic code inscribed in the DNA is transformed into proteins. The process consists of four main steps:   \emph{transcription} of a DNA gene into an mRNA molecule,    \emph{translation} of the mRNA molecule  to a protein,  degradation of mRNA molecules, and   degradation of proteins. During mRNA translation, macromolecules called ribosomes move unidirectionally along the mRNA molecule, decoding it codon by codon into a corresponding chain of amino acids that is folded to become a functional protein. Translation is a fundamental biological process, and understanding and re-engineering
 this process is important in  many scientific disciplines including medicine, evolutionary biology, and synthetic biology~\cite{Alberts2002}.

New methods that measure gene-specific translation activity at the whole-genome scale, like
polysome profiling \cite{Arava2003} and ribosome profiling
\cite{Ingolia2009}, have led to a growing interest  in mathematical models for translation.
Such models can be used to integrate and explain the rapidly accumulating
biological data as well as  to predict the outcome of various manipulations of the genetic machinery.
Recent methods that allow  \emph{real-time imaging} of  translation on a  single mRNA transcript in 
vivo (see, e.g.~ \cite{Yan2016,Wu2016,Morisaki2016,ChongWang2016}) are expected to provide even more motivation for developing and analyzing powerful dynamical models of translation.

Down-regulation of    translation is important in cell biology, medicine, and biotechnology.
For example, in many organisms small RNA genes, such as microRNAs, hybridize to the mRNA in specific locations~\cite{Ghildiyal2009,Inui2010} in order to
down-regulate translation initiation or elongation \cite{Fabian2010,Filipowicz2008} and/or promote mRNA  degradation.  Alterations in the expression of microRNA  genes contribute to the pathogenesis of most, if not all,  human malignancies \cite{Croce2009}, and many times cancer cells are targeted via generating tumor specific RNA interference (RNAi)  genes that down-regulate the oncogenes~\cite{Tavazoie2008,Zhang2003,Devi2006}. Furthermore, many viral therapeutic treatments
 and viral vaccines are based on the attenuation of mRNA translation in the viral genes \cite{Ben-Yehezkel2015,Goz2015,Kaspar2005,Coleman2008,Perez2009}.
Down regulation of mRNA translation in an \emph{optimal} manner is also related to fundamental biomedical topics such as molecular evolution and functional genomics \cite{Tuller2010c,Forman2008,Fang2004}.

Here we study  for the first  time  optimal down regulation of translation in  a dynamical  model of translation.
A standard model for translation is the \emph{totally asymmetric simple exclusion process}~(TASEP) \cite{Shaw2003,TASEP_tutorial_2011}. In this model, particles  hop randomly  along an ordered lattice of sites. Simple exclusion means that a particle cannot hop into a site that is occupied by another particle. This models hard  exclusion between the particles, and creates an indirect
 coupling between the particles.  
Indeed, if a particle remains in the same site for a long time then all the particles 
preceding  this site cannot move forward leading to   a ``traffic jam''. 

In the context of translation, the lattice is the mRNA molecule; the particles are the ribosomes; and hard  exclusion means that a ribosome cannot move forward if the codon in front of it is covered by another ribosome.
In the \emph{homogeneous} TASEP~(HTASEP) all the transition rates within the lattice are assumed to be equal and normalized to $1$, and thus the model is specified by an input rate $\alpha$, an exit rate $\beta$, and an order $N$ denoting the number of sites in the lattice.
TASEP is a fundamental model in non-equilibrium statistical mechanics that has been used to model numerous natural and artificial  processes including traffic flow, surface growth, communication networks, evacuation dynamics
and more~\cite{TASEP_book,tasep_ad_hoc_nets}.

\begin{figure}
 \begin{center}
\includegraphics[width= 14cm,height=3cm]{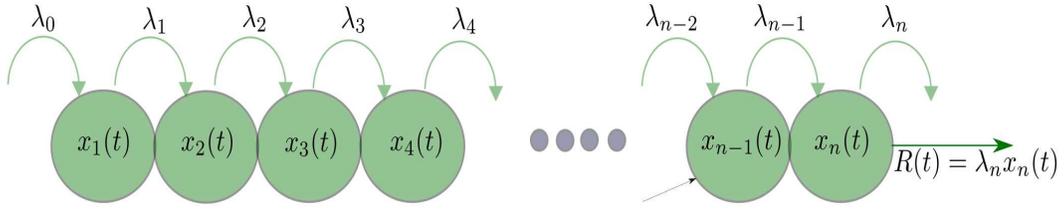}
\caption{The RFM models unidirectional flow along a chain of $n$ sites.
The state variable~$x_i(t)\in[0,1]$ represents
 the density of site $i$ at time $t$. The parameter $\lambda_i>0$ controls the transition rate from  site~$i$ to site~$i+1$, with~$\lambda_0>0$ [$\lambda_n>0$] controlling    the initiation [exit] rate. The output rate at time $t$ is~$R(t) =\lambda_n x_n(t)$.  } \label{fig:rfm}
\end{center}
\end{figure}

The \emph{ribosome flow model}~(RFM)~\cite{reuveni} is a nonlinear, continuous-time
 compartmental model for the unidirectional  flow of ``material" along a chain of $n$ consecutive compartments (or sites).  It can be derived via  a
 mean-field approximation of TASEP~\cite{TASEP_book,solvers_guide}.
In the~RFM, the state variable $x_i(t): \R_+ \to [0,1]$, $i=1,\dots,n$, describes the normalized amount (or density)
of ``material''   in site~$i$ at time~$t$, where $x_i(t)=1$ [$x_i(t)=0$]
 indicates that site $i$ is completely full [completely empty] at time $t$.
Thus, the vector~$x(t):=\begin{bmatrix}x_1(t)&\dots&x_n(t)\end{bmatrix}'$ describes the
 density profile  along the chain at time~$t$.  A   parameter~$\lambda_i>0$, $i=0,\dots,n$, controls the transition rate from site~$i$ to site~$i+1$, where~$\lambda_0$ [$\lambda_n$] is the initiation [exit] rate (see Fig.~\ref{fig:rfm}).
 The output rate at time~$t$ is~$R(t)=\lambda_n x_n(t)$.
In the context of translation, the  ``material'' are the moving  ribosomes, and each site represents a group of codons, i.e. the mRNA is coarse-grained into~$n$ consecutive sites of codons. Thus,~$R(t)$, the output flow of ribosomes at time~$t$, is    the \emph{protein production rate} at time $t$. It is known that the RFM admits a unique \emph{steady-state} production rate denoted by $R=R(\lambda)$~\cite{RFM_stability}, where~$\lambda:=\begin{bmatrix}\lambda_0&\dots\lambda_n\end{bmatrix}'$.

\begin{figure}
 \begin{center}
\includegraphics[width= 8cm,height=7cm]{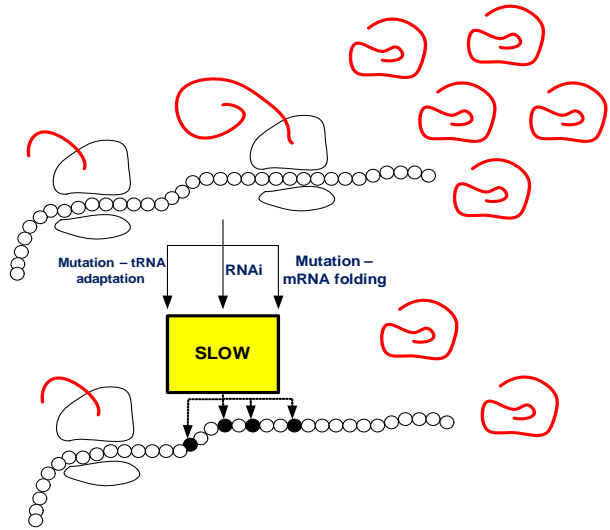}
\caption{ The problem we consider is how  to efficiently select transition rates along the mRNA molecule, within a given set of possible rates, such that the protein production rate is minimized. In practice, translation rate modification  can be done by introducing mutations into the gene or by designing a corresponding RNAi molecule. } \label{fig:Figure_RFM_blocking}
\end{center}
\end{figure}

Here, we use the RFM to  analyze how to maximally down-regulate  mRNA translation.
To do this, we formulate  the following general optimization problem. Given an  mRNA molecule with~$n$ sites, and a convex and compact region of feasible transition rates $\Omega^{n+1}$, find a vector~$\lambda^* \in\Omega^{n+1} $ such that~$R(\lambda^*)=\min_{\lambda\in\Omega^{n+1} }R(\lambda)$. In other words, the problem is how to select transition rates, within a feasible region, such that the production rate is minimized  (see Fig.~\ref{fig:Figure_RFM_blocking}).  To the best of our knowledge, this is the first time that such a problem is analyzed in a dynamical  model of mRNA translation.

As a concrete  example, consider an RFM with dimension $n$ and   rates~$\BLMD$. Given  a ``total reduction budget''~$b\in[0,\min\{\bar \lambda_i\}]$, define the feasible set~$\Omega^{n+1}\subset \R^{n+1}_+$   by
	\[
	      \left \{ \begin{bmatrix} \bar \lambda_0-\varepsilon_0 &\dots& \bar \lambda_n-\varepsilon_n  \end{bmatrix} :\varepsilon_i\geq 0,  \;\varepsilon_0+\dots+\varepsilon_n=b  \right   \}.
	\]
In other words, the feasible set is the set of all the
 rates obtained by applying a ``total reduction budget''~$b$ in  the rates of the given  mRNA molecule. The question is how to distribute the total reduction budget over the rates so as to obtain the minimal  possible protein production rate.
 We prove  that:
\begin{itemize}
\item If some rate~$\bar \lambda_k$ is a ``bottleneck" rate,
in a sense that will be made precise below, then an  optimal reduction in protein production rate is obtained by
				using all the reduction budget~$b$ to further decrease~$\bar \lambda_k$;
\item If all the given rates are equal, i.e.~$\bar \lambda_0=\dots=\bar \lambda_n$, then the transition rate
at the middle of the mRNA molecule is the bottleneck rate,  and thus
an optimal reduction in protein production rate is obtained by using all the reduction budget to reduce this transition rate.
\end{itemize}
Thus, in this case there exists a single site such that   mutating it yields  the maximal inhibition
of translation.   Our results allow to determine where this site is located.


The remainder of this paper is organized as follows. We first 
 briefly review  some known results on  the RFM that are needed for our purposes.
The following section  poses the problem of  down-regulating
 the steady-state protein production rate in the RFM in an optimal manner,
 and then describes our main results. Analysis of the RFM is non-trivial, as this is a nonlinear dynamical model.
In particular, the mapping from~$\lambda$ to~$R(\lambda)$
is nonlinear and does not admit  a closed-form expression. Nevertheless,
by combining  tools from convex optimization
 and eigenvalue sensitivity theory, we show that this optimization problem is tractable in some cases, and
rigorously prove several results
 that have interesting biological implications.
The final section summarizes and describes
several directions for further research. To increase the readability of this paper, all the proofs are placed in the Appendix.

 \section*{Ribosome Flow Model}

 The dynamics of the RFM with $n$ sites is given by $n$ nonlinear first-order ordinary differential equations:
\begin{align}\label{eq:rfm}
                    \dot{x}_1&=\lambda_0 (1-x_1) -\lambda_1 x_1(1-x_2), \nonumber \\
                    \dot{x}_2&=\lambda_{1} x_{1} (1-x_{2}) -\lambda_{2} x_{2} (1-x_3) , \nonumber \\
                    \dot{x}_3&=\lambda_{2} x_{ 2} (1-x_{3}) -\lambda_{3} x_{3} (1-x_4) , \nonumber \\
                             &\vdots \nonumber \\
                    \dot{x}_{n-1}&=\lambda_{n-2} x_{n-2} (1-x_{n-1}) -\lambda_{n-1} x_{n-1} (1-x_n), \nonumber \\
                    \dot{x}_n&=\lambda_{n-1}x_{n-1} (1-x_n) -\lambda_n x_n.
\end{align}
If we define~$x_0(t):=1$ and $x_{n+1}(t):=0$
then~\eqref{eq:rfm} can be written more succinctly as
\be\label{eq:rfm_all}
\dot{x}_i=\lambda_{i-1}x_{i-1}(1-x_i)-\lambda_i x_i(1-x_{i+1}),\quad i=1,\dots,n.
\ee
Eq.~\eqref{eq:rfm_all}  can be explained as follows. The flow of material from site~$i$ to site~$i+1$ at time~$t$
is~$\lambda_{i} x_{i}(t)(1 - x_{i+1}(t) )$. This flow is proportional to $x_i(t)$, i.e. it increases with the density at site~$i$, and to $(1-x_{i+1}(t))$, i.e. it decreases as site~$i+1$ becomes fuller.  This corresponds to a ``soft''  version of a simple exclusion principle. Note that the maximal possible  flow  from site~$i$ to site~$i+1$  is the transition rate~$\lambda_i$.


Let~$x(t,a)$ denote the solution of~\eqref{eq:rfm}
at time~$t \ge 0$ for the initial
condition~$x(0)=a$. Since the  state-variables correspond to normalized density levels, with~$x_i(t)=0$ [$x_i(t)=1$] representing that site~$i$
is completely empty [full] at time~$t$,
  we always assume that~$a$ belongs to the  closed $n$-dimensional
  unit cube:
$
           C^n:=\{x \in \R^n: x_i \in [0,1] , i=1,\dots,n\}.
$
Let $\Int(C^n)$ [$\partial C^n$] denote the interior [boundary] of $C^n$. It  is straightforward to verify that~$\partial C^n$ is repelling, i.e. if $a\in\partial C^n$ then $x(t,a)\in\Int(C^n)$ for all $t>0$, so~$C^n$ and also~$\Int(C^n)$ are invariant
sets for the dynamics.

An important property of the~RFM is the symmetry between the ``particles'' (i.e. ribosomes) moving from left to right and ``holes'' (i.e. "lack" of ribosomes)
  moving from right to left (in the
	TASEP literature,
	this property is sometimes referred to as the ``particle-hole" symmetry). Indeed,  let~$q_j(t):=1-x_{n+1-j}(t)$, $i=1,\dots,n$.
Then
\begin{align*}
										\dot q_1&= \lambda_n(1-q_1) -\lambda_{n-1}q_1(1-q_2),      \\
									\dot q_2&= \lambda_{n-1}q_1(1-q_2) -\lambda_{n-2}q_2(1-q_3)  ,    \\
									&\vdots\\
									\dot q_n&= \lambda_{1}q_{n-1}(1-q_n) -\lambda_{0}q_n.
\end{align*}
This is another  RFM, but now with rates $\lambda_n,\dots,\lambda_0$.

The RFM has been used  to model and analyze
the flow of ribosomes along the mRNA molecule during the process of mRNA translation.
The (soft) simple  exclusion principle  corresponds to the fact that ribosomes have volume and cannot overtake one another.

It is important to mention that it has been shown in \cite{reuveni} that the correlation between the production rate based on modeling using RFM and using TASEP over all {\em S. cerevisiae} endogenous genes is~$0.96$. In addition, it has also been shown there that the RFM model agrees well with biological measurements of ribosome densities. Furthermore, it was also shown that the RFM model predictions correlate well (correlations up to~$0.6$) with protein levels in various organisms (e.g. {\em E. coli}, {\em S. pombe}, {\em S. cerevisiae}).
Given the high levels of bias and
noise  in measurements related to gene expression and the inherent
stochasticity  of intracellular biological processes  (see e.g. \cite{Diament2016,Kaern2005}), these  correlation values   demonstrate the relevance of the RFM in this context.

\subsection{Steady-State Spectral Representation}
Ref.~\cite{RFM_stability} has shown that the RFM is a
\emph{tridiagonal cooperative dynamical system}~\cite{hlsmith},
and    that~\eqref{eq:rfm}
admits a \emph{unique} steady-state point~$e=e(\LMD) \in \Int(C^n)$ that is globally asymptotically stable, that is, $\lim_{t\to\infty} x(t,a)=e$  for all $a\in C^n$ (see also~\cite{RFM_entrain}). This means that the ribosomal
density profile always converges to a steady-state profile that depends on the rates, but not  on the initial condition. In particular, the output rate~$R(t)=\lambda_n x_n(t)$
 converges to a steady-state   value~$R:=\lambda_n  {e}_n$.

At steady-state (i.e, for~$x=e$), the left-hand side of all the equations
in~\eqref{eq:rfm} is zero, so
\begin{align} \label{eq:ep}
                      \lambda_0 (1- {e}_1) & = \lambda_1 {e}_1(1- {e}_2)\nonumber \\&
                      = \lambda_2  {e}_2(1- {e}_3)   \nonumber \\ & \vdots \nonumber \\
                    &= \lambda_{n-1} {e}_{n-1} (1- {e}_n) \nonumber \\& =\lambda_n  {e}_n\nonumber\\&=R.
\end{align}
This yields
\begin{align}\label{eq:rall}
R=\lambda_i e_i(1-e_{i+1}), \quad i=0,\dots,n,
\end{align}
where $e_0:=1$ and $e_{n+1}:=0$.
Ref.~\cite{RFM_max} used these expressions to
 provide  a  \emph{spectral representation} of the mapping from the set of rates~$\lambda$  to the steady-state output rate~$R$.
Let
$\R^n_+:=\{y\in \R^n: y_i \geq 0,\; i=1,\dots,n\}$ and
$\R^n_{++}:=\{y\in \R^n: y_i> 0,\; i=1,\dots,n\}$.

\begin{Theorem}\cite{RFM_max}\label{thm:spect}
Given an RFM with rates~$\lambda=\begin{bmatrix}\lambda_0&\dots&\lambda_n \end{bmatrix}'$, let~$R=R(\lambda)$
denote its steady-state production rate. Define
 an $(n+2)\times(n+2)$ Jacobi  matrix~$A=A(\lambda)$ by
\be\label{eq:bmatrox}
                  A:=  \begin{bmatrix}
 0 &  \lambda_0^{-1/2}   & 0 &0 & \dots &0&0 \\
\lambda_0^{-1/2} & 0  & \lambda_1^{-1/2}   & 0  & \dots &0&0 \\
 0& \lambda_1^{-1/2} & 0 &  \lambda_2^{-1/2}    & \dots &0&0 \\
 & &&\vdots \\
 0& 0 & 0 & \dots &\lambda_{n-1}^{-1/2}  & 0& \lambda_{n }^{-1/2}     \\
 0& 0 & 0 & \dots &0 & \lambda_{n }^{-1/2}  & 0
  \end{bmatrix} .
\ee
Then:
\begin{enumerate}
\item The eigenvalues of~$A$ are real and distinct, and if we order them as~$\zeta_1<\dots<\zeta_{n+2}$
then~$\zeta_{n+2}=(R(\lambda))^{-1/2}$.
\item Let~$s_i(\lambda) :=\frac{\partial }{\partial \lambda_i} R(\lambda)$, i.e. the sensitivity of~$R$ with respect to (w.r.t.) the rate~$\lambda_i$.
Let $v\in\R^{n+2}_{++}$ denote an  eigenvector of~$A$
corresponding to the eigenvalue $\zeta_{n+2}$. Then
\be\label{eq:sense}
s_i(\lambda)  = \frac{2 R^{3/2}}{\lambda_i^{3/2} v'v} v_{i+1} v_{i+2} ,\quad i=0,\dots,n.
\ee
\end{enumerate}
\end{Theorem}

This means that the steady-state production rate, and its sensitivity with respect to the transition rates,
 can be computed efficiently using numerical algorithms for computing the
eigenvalues and eigenvectors of tridiagonal matrices.
 Theorem~\ref{thm:spect} also implies that
\be\label{eq:rhom}
R(c\lambda_0,\dots,c\lambda_n)=cR( \lambda_0,\dots, \lambda_n),\quad\text{for all }c>0,
\ee
i.e. $R(\lambda)$ is homogeneous of degree one.

Another  important implication of Theorem~\ref{thm:spect}
is that~$R$ is a \emph{strictly concave} function of the transition rates~$\{\lambda_0,\dots,\lambda_n\}$ over~$\R^{n+1}_{++}$~\cite{RFM_max}. Also, it implies that~$\frac{\partial }{\partial \lambda_i} R>0$ for all~$i$,
that is, an increase in any of the rates yields an increase
in the steady-state production rate.

For more on the analysis of the RFM, and also networks of interconnected~RFMs,
 using tools from systems and control theory, see e.g.~\cite{zarai_infi,RFM_max,RFM_sense,RFM_feedback,RFMR,rfm_control,RFM_model_compete_J}.

\section*{Main Results }\label{sec:main}
We  begin by posing  a general minimization problem for the steady-state production rate in the~RFM.

\begin{Problem}\label{prob:min}
Given a convex and compact feasible set of transition rates~$\Omega^{n+1} \subset  \R^{n+1}_{++}$, find
$\lambda^*\in\Omega^{n+1}$ such that~$R(\lambda^*)=
				\min_{\lambda\in\Omega^{n+1}}R(\lambda).
$
\end{Problem}

From the biological point of view, the
 feasible set of transition rates~$\Omega^{n+1}$ depends
on  all the biophysical constraints
 on the transition  rates along the coding sequence. For example,   the maximal/minimal decoding rate of a codon
 (e.g. via its adaptation to the tRNA pool) \cite{Dana2014B}, the maximal possible effect of mRNA folding (after codon substitution) on each codon \cite{TullerGB2011}, the maximal possible effect (after amino acid substitution) of the interaction of the ribosome with amino acids of the nascent peptide~\cite{Sabi2015}, and the maximal elongation slow down due to interaction with microRNAs~\cite{Ghildiyal2009,Inui2010}.

Below we explain how to pose
 various interesting biological problems in the framework of  Problem~\ref{prob:min}.
Examples include finding
   the minimal number of mutations that down regulate translation of a gene/mRNA under a certain ``total reduction budget''.  This is practically  important when we use costly (in terms of time and money) gene editing approaches.
Another related question is how to down regulate translation of a gene/mRNA with
 a maximal number of mutations. This is     important when   attenuating viral replication rate for generating a safe live attenuated vaccine. A large number of  mutations reduces the probability of reverting.
One may also define the feasible set in Problem~\ref{prob:min} in such a way that some rates cannot be changed.
This is relevant for example when some  codons along the mRNA cannot be modified. Indeed,  various positions along the mRNA   affect regulatory mechanisms that we may not want to alter (e.g. co-translational folding, splicing, translation).

It is well-known  (see, e.g.~\cite[Thm.~7.42]{beck2014})
that if~$f:\Omega^{n+1} \to\R$ is a continuous   and  {strictly} convex function defined
over a convex and compact set~$\Omega^{n+1}$
then all the maximizers of~$f$ over~$\Omega^{n+1}$ are extreme points of~$\Omega^{n+1}$ (for more on the problem of maximizing a convex function, or equivalently, minimizing a concave  function, see e.g.~\cite{concave_prog}).
Combining this with the fact that~$R$ is a strictly concave function of the transition
rates over~$\R^{n+1}_{++}$ implies the following.
\begin{Proposition}\label{prop:main_sol}
												Every solution
							   of Problem~\ref{prob:min} is
								an extreme point  of~$\Omega^{n+1}$.
\end{Proposition}
In particular, if the set of extreme points of~$\Omega^{n+1}$ is finite then
one can always solve Problem~\ref{prob:min} by simply
calculating~$R(\lambda)$ for all~$\lambda$ that
are extreme   points of~$\Omega^{n+1}$, and then
finding  the minimum    of these values.
In particular, if~$\Omega^{n+1}$ is a convex
polytope  then the extreme points are just the vertices of~$\Omega^{n+1}$.
Thus, when the biophysical constraints lead to a feasible set of rates that is a
convex polytope then it is
  computationally  straightforward to
	determine how to modify the rates so as to
	obtain the largest decrease in translation rate under reasonable biophysical constraints.

In the remainder of this section, we consider three
special cases of Problem~\ref{prob:min} for which it is also possible
to obtain analytic results.

\begin{Problem}\label{prob:sd}
Given an  RFM  with $n$ sites,  rates~$\bar \lambda_0,\dots, \bar \lambda_n$,
  and a ``total reduction  	budget''~$b\in[0,\min\{\bar \lambda_i\}]$,
	let~$\Omega^{n+1}=\Omega^{n+1}(\bar \lambda,b) $ be the set
	\be\label{eq:omeg_cov}
	  \left \{\begin{bmatrix} \bar \lambda_0-\varepsilon_0,\dots,\bar \lambda_n-\varepsilon_n\end{bmatrix} : \varepsilon_i \geq 0,
	\sum_{i=0}^n \varepsilon_i=b \right \}.
	\ee
	Find~$\lambda^* \in \Omega^{n+1} $ such that~$R(\lambda^*)=\min_{\lambda\in\Omega^{n+1} } R(\lambda)$.
	\end{Problem}
In other words, $\Omega^{n+1} $ is the set of
	all the rates that can be obtained by applying a total reduction~$b$ to  the given rates~$\bar \lambda_i$.
	From a mathematical point of view, $b$ provides a bound on the total possible rate reduction.
	It also couples the reduction in different rates, as a larger reduction in one rate
	must be compensated  by smaller reductions in other rates so that the total reduction will not exceed~$b$.
	From a synthetic biology point of view,~$b$ can be used to capture the idea
of maximally inhibiting the production rate  while
 minimizing the side-effects  of this down regulation. For example, a very small value of~$b$
forces a solution with small modifications in all the rates. This is expected of course
to minimize the
effect of the mutations on  the   fitness of the cell/organism.
 For example, since co-translation folding \cite{Zhang2009,Pechmann2013,Tuller2015} is related to the ribosome
transition rates along the mRNA, smaller changes in the rates are expected to have a
 smaller effect on protein folding (and thus on the functionality of the protein and the overall
organismal fitness).
Smaller changes in the transition rates are also related to a ``simpler'' biological
solution in the sense of fewer mutations, less miRNAs, etc.

The next example demonstrates Problem~\ref{prob:sd}.
\begin{Example}\label{exa:firdem}
Consider an RFM with dimension $n=4$ and   transition rates
\[
\bar \lambda_0=0.85,\; \bar \lambda_1=0.92, \;\bar \lambda_2=0.78,\;\bar \lambda_3=0.57, \; \bar \lambda_4=0.88.
\]
The steady-state production rate is $R(\bar \lambda_0,\dots,\bar \lambda_4) = 0.2308$ (all numbers are to four digit accuracy). Suppose  that the total reduction budget is $b=0.1$.
Then, for example, the vector
\[
\lambda:=\bar \lambda-\begin{bmatrix} 0.05 & 0 & 0 & 0.02 & 0.03 \end{bmatrix},
\]
belongs to~$\Omega^5$, and~$R(\lambda)=0.2260$.
An  optimal solution is  $\lambda^*:=\begin{bmatrix} 0.85 & 0.92 & 0.78 & 0.47 & 0.88 \end{bmatrix}'\in\Omega^5$, with $R(\lambda^*)=0.2140$.
Note that this corresponds to reducing~$b$ from the rate~$\bar \lambda_3$, which is the minimum
of all the rates~$\bar \lambda_i$, leaving all the other rates unchanged. \hfill{$\square$}
\end{Example}

Let~$d^i\in\R^{n+1}$ denote the~$(i+1)$'th  column of the~$(n+1)\times(n+1)$ identity matrix.
The set~$\Omega^{n+1}(\bar \lambda,b)$ is a convex polytope with vertices:
\[
			 v^i : = \begin{bmatrix} \bar \lambda_0&\dots&\bar \lambda_n \end{bmatrix}' -b  d^i ,\quad i=0,\dots,n.
\]

If there exists an index~$i$ such that~$\bar \lambda_i=b$ then it is clear that an optimal solution is to reduce~$\bar \lambda_i$ to~$0$, as then the steady-state production rate will be zero. So  we always assume that~$b$ takes values in the set~$[0,\min\{\bar \lambda_i\}-\rho]$, for some~$\rho>0$. This means that Problem~\ref{prob:sd} is a special case of Problem~\ref{prob:min}, as
 $\Omega^{n+1}(\bar \lambda,b)$ is a convex polytope contained in~$\R^{n+1}_{++}$.

 By Prop.~\ref{prop:main_sol}, every solution of Problem~\ref{prob:sd} is  contained in the set~$\{v^0,\dots, v^n\}$.
  In other words, every
  minimizer  corresponds to
reducing \emph{all} the available  budget~$b$ from a single  rate.
This immediately yields a simple and efficient algorithm for solving Problem~\ref{prob:sd}:
use the spectral representation of~$R$ to compute~$R(v^i)$, $i=0,\dots,n$, and then find the minimum
of all these values. Since the matrix~$A$ in~\eqref{eq:bmatrox} is symmetric and tridiagonal, calculating~$R(v^i)$ can be done
efficiently even for large values of~$n$.
We wrote a simple  (and unoptimized) MATLAB script for
solving Problem~\ref{prob:sd}, and ran it on a MAC laptop
with a~$2.6$ GHz Intel core $i7$ processor.
For an RFM with  $n=500$ (a typical coding region  includes
 a few  hundred  codons \cite{Zhang2000}), rates $\bar \lambda_i=1$, $i=0,\dots,500$, and~$b=0.1$,
the optimal solution is found in~$3.14$ seconds.

Example~\ref{exa:firdem} may suggest that reducing the slowest transition
 rate by~$b$ always  yields an  optimal solution, but in general this is not true
(see Example \ref{ex:bn} below).

One  may also consider a different
  feasible set in Problem~\ref{prob:sd}, namely,
\[
 \left \{\begin{bmatrix} \bar \lambda_0-\varepsilon_0,\dots,\bar \lambda_n-\varepsilon_n\end{bmatrix} : \varepsilon_i \geq 0,
	\sum_{i=0}^n \varepsilon_i\leq b \right \} ,
	\]
i.e. here the total reduction is \emph{up to}~$b$.
However, by Theorem~\ref{thm:spect} $\frac{\partial }{\partial \lambda_i}R(\lambda) >0$ for all $i$, and thus an
optimal solution for this problem is guaranteed to agree with an optimal solution of
 Problem~\ref{prob:sd}.

The next example demonstrates the effect of increasing the total reduction rate~$b$
on the optimal solution of Problem~\ref{prob:sd}.

\begin{figure}
 \begin{center}
\includegraphics[width= 8cm,height=7cm]{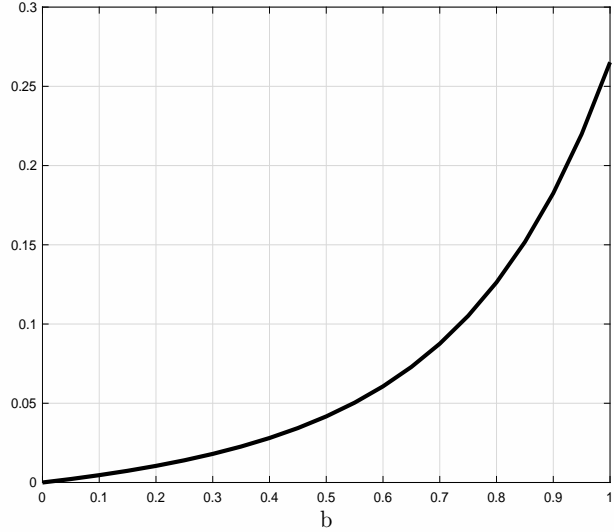}
\caption{$\Delta R$ as a function of $b$ for an RFM with dimension $n=10$ and rates~$\bar \lambda_i=1$, $i=0,\dots,10$.} \label{fig:n10_dR}
\end{center}
\end{figure}

\begin{Example}
Consider an RFM with dimension $n=10$, and rates~$\bar \lambda_i=1$, $i=0,\dots,n$. Here $R(\bar \lambda)=0.2652$.
We calculated the optimal solution~$\lambda^*$ for different values of~$b$, and also the value
 $\Delta R(b):=R(\bar \lambda)-R(\lambda^*)$, that is,
the     optimal reduction in protein rate that can be obtained for various
values of~$b$.
 Figure~\ref{fig:n10_dR} depicts~$\Delta R$ as a function of~$b$.   It may be seen that~$\Delta R$ increases quickly
 with~$b$ (specifically, the relation is superlinear).\hfill{$\square$}

\end{Example}

\subsection{Optimal reduction and sensitivities}
It is also possible to derive theoretical results on the structure of an optimal solution~$\lambda^*$ in Problem~\ref{prob:sd}
using the sensitivities $s_i(\lambda) :=\frac{\partial}{\partial \lambda_i} R(\lambda)$.
Note that these can be computed efficiently using~\eqref{eq:sense}.
\begin{Proposition}\label{prop:sense}
Consider   Problem~\ref{prob:sd}.
If there exist~$i,j\in\{0,\dots,n\}$ such that
\be\label{eq:sencom}
				s_i(\bar \lambda)<s_j(\bar \lambda )
\ee
then any  optimal solution~$\lambda^*$ satisfies~$\lambda^*_i=\bar \lambda_i$.
\end{Proposition}
In other words, if the sensitivity of the steady-state production rate to rate~$\lambda_i$ at~$\bar \lambda$ is  lower than some other sensitivity
then an optimal solution will \emph{not} include a reduction in~$\bar \lambda_i$. Indeed, it is better to
distribute the reduction budget over some other, more sensitive,  rates.

\begin{Remark}\label{rem:sens}
Note that since $R $ is a strictly concave function of the rates,
\begin{align*}
\frac{\partial }{\partial \lambda_i}s_i(\bar\lambda)
&=\frac{\partial^2}{\partial \lambda_i^2} R(\bar \lambda)
 <0,
\end{align*}
for any~$\bar \lambda\in\R^{n+1}_{++}$ and any~$ i\in\{0,\dots,n\}$.
In other words, a  decrease in $\bar \lambda_i$ increases the sensitivity w.r.t.   this rate.
\end{Remark}

Proposition~\ref{prop:sense}  leads to the following definition.
\begin{Definition}\label{def:btl}
Given an RFM with rates~$\bar \lambda$, a
  transition rate $\bar \lambda_j$ is called a \emph{bottleneck  rate}
	if $s_j(\bar \lambda) >  s_i(\bar \lambda)$, for all $i\ne j $.
\end{Definition}
In other words, a bottleneck  rate is one with a maximal sensitivity.

Combining this with Proposition~\ref{prop:sense} immediately yields the following result.
\begin{Corollary}\label{coro:optsimp}
Given an RFM with rates~$\bar \lambda$, suppose that~$s_j(\bar\lambda )$ is a   bottleneck  rate.
Then  the unique optimal solution to Problem~\ref{prob:sd} is obtained by reducing~$\bar \lambda_j$
by~$b$.
\end{Corollary}

An important observation is that the slowest
 rate along the~mRNA molecule and the bottleneck rate may be different.
The next example demonstrates this.

 \begin{Example}\label{ex:bn}
 Consider an RFM with dimension $n=4$, and rates~$\bar \lambda_3=1.85$, $\bar \lambda_i=2.0$, $i=0,1,2,4$. In this case,
$s_0(\bar \lambda)=0.0297 $, $s_1(\bar \lambda)=0.0687$,
 $s_2(\bar \lambda)=0.0901$, $s_3(\bar \lambda)= 0.0856$,
and~$s_4(\bar \lambda)=0.0343 $.
 Thus, although
the minimal rate is~$\bar \lambda_3$,  the bottleneck rate is~$\bar \lambda_2$.
In particular, the optimal solution will be to reduce~$\bar \lambda_2$ by~$b$, and not~$\bar \lambda_3$, even though~$\bar \lambda_3$
is the minimal rate.\hfill{$\square$}
 \end{Example}
However, note that Remark~\ref{rem:sens} implies that
if some rate~$\lambda_i$ is decreased enough then it will eventually  become
a bottleneck rate.

Proposition~\ref{prop:sense} can be used to derive analytic results in cases
  where we can obtain   explicit information on
the sensitivities at a point~$\bar\lambda \in \R^{n+1}_+$. The next two results demonstrate  this.

\begin{Proposition}\label{prop:cases}
Consider an RFM with dimension~$n$ and with equal rates, i.e.~$\bar \lambda_0=\dots=\bar \lambda_n$. If~$n$ is even then the unique  optimal solution to
 Problem~\ref{prob:sd} is:~$\lambda^*=\bar \lambda-bd^{  n/2  }$.
If~$n$ is odd then there are two optimal solutions:~$\lambda^*=\bar \lambda-bd^{\lfloor n/2 \rfloor}$
and~$\lambda^*=\bar \lambda-bd^{\lfloor n/2 \rfloor+1}$.
\end{Proposition}

In other words, in the case where all the   rates are equal, the bottleneck is  at the center of the chain.
These results are closely related to the fact that in a dynamic model for phosphorelay~\cite{phos_relays}, that is very similar to the~RFM,
the middle layer in the model is the most sensitive to changes in the input. 
This also  agrees with the so called ``edge-effect''  in the HTASEP~\cite{toward_prod_rates,edge_tasep_2009,PhysRevE.76.051113}, i.e. the fact that the steady-state output rate is less sensitive to the rates that are
close to the edges of the chain.
For more on the sensitivity of TASEP to manipulations in  the
initiation, hopping, and exit rates, see~\cite{PhysRevE.76.051113,foulaadvand2008asymmetric,Chou2004,PhysRevE.58.1911}.

Another case where analytic results can be derived is when the rates
in the RFM lead to equal steady-state occupancies along the mRNA molecule.
This happens when $\lambda_1=\lambda_2=\cdots=\lambda_{n-1}=\lambda_0+\lambda_n$
(see~\eqref{eq:ep}).

\begin{Proposition}\label{prop:e_uniform}
Consider an RFM with dimension~$n$ and rates~$\bar \lambda$ such that
 $\bar e_1=\cdots=\bar e_n:=e_c$, i.e. all the steady-state occupancies are equal, and $e_c$ denotes their common value.

\begin{enumerate}
\item If $e_c < 1/2$ then the unique  optimal
solution to Problem~\ref{prob:sd} is
\be\label{eq:opt_un1}
\lambda^*= \bar \lambda - bd^0.
\ee
\item If $e_c > 1/2$ then the unique  optimal
solution to Problem~\ref{prob:sd} is
\be\label{eq:opt_un2}
\lambda^*= \bar \lambda - bd^n.
\ee
\item If $e_c=1/2$ then~\eqref{eq:opt_un1}  and~\eqref{eq:opt_un2} are the optimal solutions.
\end{enumerate}
\end{Proposition}

In other words, if the equal occupancy is relatively low [high] then
maximal inhibition of the production rate is obtained by reducing the total reduction rate
from the initiation [exit] rate, leaving all the other rates unchanged.
.

\begin{Example}
Consider   Problem~\ref{prob:sd} for an RFM with~$n=5$,
rates~$\bar\lambda=\begin{bmatrix} 1 & 5/2 & 5/2 & 5/2 & 5/2 & 3/2 \end{bmatrix}'$, and~$b=1/2$. Note that in this case~$\bar e_1=\cdots=\bar e_5=2/5$.
A calculation yields
$R(\bar \lambda - bd^0)=0.3999$,
$R(\bar \lambda - bd^1)=0.5651$,
$R(\bar \lambda - bd^2)=0.5762$,
$R(\bar \lambda - bd^3)=0.5829$,
$R(\bar \lambda - bd^4)=0.5874$,
and
$R(\bar \lambda - bd^5)=0.5746$,
so the optimal solution is~$\lambda^*=\bar \lambda - bd^0$.
Since
  $e_c<1/2$, this agrees with Proposition~\ref{prop:e_uniform}.\hfill{$\square$}
\end{Example}

In some cases it may be more natural to define the transition rate reduction in relative rather than
 absolute terms. This is captured by the  following optimization problem.

\begin{Problem}\label{prob:sd_rel}
Given an  RFM  with $n$ sites,  rates~$\bar \lambda_0,\dots, \bar \lambda_n$,
  and a  total reduction budget~$q\in[0,1)$,
	let~$\Gamma^{n+1}=\Gamma^{n+1}(\bar \lambda,q) \subset \R^{n+1}_{++}$ be the set
	\be\label{eq:omeg_cov_rel}
	  \left \{\begin{bmatrix} \bar \lambda_0(1-\delta_0),\dots,\bar \lambda_n(1-\delta_n)\end{bmatrix} : \delta_i \geq 0,
	\sum_{i=0}^n \delta_i=q \right \}.
	\ee
	Find~$\lambda^* \in \Gamma^{n+1} $ such that~$R(\lambda^*)=\min_{\lambda\in\Gamma^{n+1} } R(\lambda)$.
\end{Problem}
For~$i\in \{0,\dots,n \}$,
let $D^i\in\R^{(n+1)\times(n+1)}$ denote the $(n+1)\times(n+1)$ identity matrix,
but  with entry~$(i+1,i+1)$ changed to~$1-q$. The set $\Gamma^{n+1} $ is a convex polytope with vertices
$
			 u^i : =   D^i \bar \lambda 
$, $i=0,\dots,n$. Thus, Problem~\ref{prob:sd_rel} is also a special case of Problem~\ref{prob:min}, and so the minimizer~$\lambda^*$
satisfies~$\lambda^* \in \{u^0,\dots,u^n\}$.


In practice, each codon (or coding region) admits
 a minimal and a maximal possible decoding rate. There are
 also   minimal and maximal
    values for the   initiation rate. These bounds
		  are determined  by  the biophysical properties of the transcript and   the intracellular environment.
To model this, we can modify the  optimization problems described above
to include a bound~$\ell_i$ on the maximal allowed  reduction
  of rate~$i$, for~$i=0,\dots,n$. The next problem demonstrates such a modification for Problem~\ref{prob:sd}.
	
\begin{Problem}\label{prob:sdg}
Consider an  RFM  with $n$ sites and rates~$\bar \lambda_0,\dots, \bar \lambda_n$.
Given a total   reduction budget~$b\in[0,\min\{\bar \lambda_i\}-\rho]$, for some~$\rho>0$,
and also bounds~$0<\ell_i<\bar \lambda_i$, $i=0,\dots,n$, with~$\sum_{i=0}^n \ell_i > b$,
let~$\Omega^{n+1}$ be as defined in Problem~\ref{prob:sd}, and let
\begin{align}\label{eq:sdg_const}
\Psi^{n+1}&:=\{\lambda\in\R^{n+1}_{++} : \lambda_i\in[\bar \lambda_i - \ell_i, \bar \lambda_i],\; i=0,\dots,n\}, \nonumber \\
\Phi^{n+1}&:=\Omega^{n+1} \cap \Psi^{n+1}.
\end{align}
	Find~$\lambda^* \in \Phi^{n+1}$ such that~$R(\lambda^*)=\min_{\lambda\in\Phi^{n+1}} R(\lambda)$.
\end{Problem}
In other words, the feasible set~$\Phi^{n+1}$ in Problem~\ref{prob:sdg} is the intersection of the set $\Omega^{n+1}$ (defined in Problem~\ref{prob:sd}), and the closed $(n+1)$-dimensional cube $\Psi^{n+1}$ that models   constraints on the maximal possible
reduction of each  rate.

 Since $\Phi^{n+1}$ is compact and  convex  (being the intersection of two compact and convex sets), Problem~\ref{prob:sdg} admits a solution that is an extreme point of $\Phi^{n+1}$. In general, not all the rates can be reduced by $b$, and thus an optimal solution  may include a reduction of \emph{several}  rates.

\begin{figure}
 \begin{center}
 \includegraphics[width= 9cm,height=8cm]{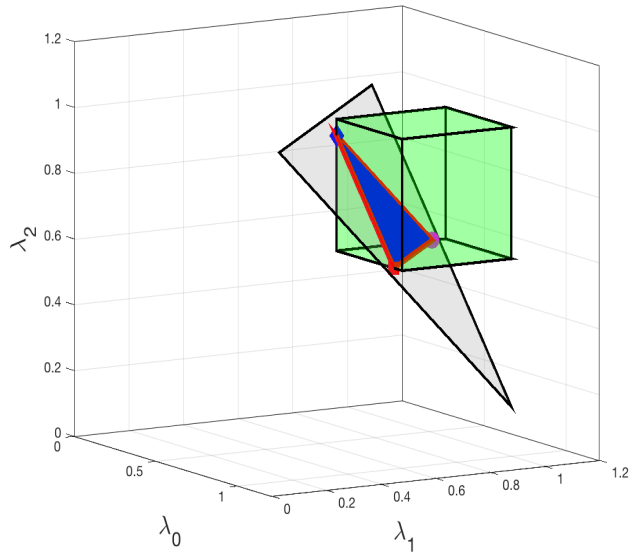}
\caption{The sets $\Omega^3$ (gray polytope), $\Psi^3$ (green cube), and $\Phi^3$ (blue polytope)   in Example~\ref{exp:sdg}.} \label{fig:poly_inter_n2}
\end{center}
\end{figure}

\begin{Example}\label{exp:sdg}
Consider Problem~\ref{prob:sdg}  for an  RFM with dimension $n=2$, rates~$\bar \lambda_i=1.0$, $i=0,1,2$, and parameters
 $b=0.85$, and $\ell_i=0.4$, $i=0,1,2$. In other words, the total possible reduction is~$0.85$, but any rate can be reduced by no more than~$0.4$.
Fig.~\ref{fig:poly_inter_n2} depicts the feasible set~$\Phi^{3}$ (blue polytope)  that is
 the intersection of the set~$\Omega^3$ (gray polytope) and the set~$\Psi^3$ (green cube). Shown also are the three extreme points of $\Phi^3$:
\begin{align*}
v^1&:=\begin{bmatrix} 0.95 & 0.6 & 0.6 \end{bmatrix}'  \text{ (red circle) },\\
 v^2&:=\begin{bmatrix} 0.6 & 0.6 & 0.95 \end{bmatrix}' \text{ (blue circle)}, \\
v^3&:=\begin{bmatrix} 0.6 & 0.95 & 0.6 \end{bmatrix}' \text{ (magenta circle)}.
\end{align*}
 A calculation yields $R(v^1)=R(v^2)=0.2538$, whereas~$R(v^3)=0.2764$. It follows that $\lambda^*=v^1$ and $\lambda^*=v^2$ are optimal solutions.
Note that these solutions correspond to reducing several rates along the mRNA molecule.
Note also that $s(\bar\lambda)=\begin{bmatrix} 0.1056 & 0.1708 & 0.1056 \end{bmatrix}'$,   so both  optimal solutions correspond  to a maximal possible reduction in a most sensitive rate, and  a maximal possible
reduction in another most sensitive rate.\hfill{$\square$}
\end{Example}

In some cases, there may be positions along the coding region
that we cannot modify
due to their potential effect on various intracellular processes.
An important advantage of Problem~\ref{prob:sdg} is that it allows
capturing this by simply setting  some of the~$\ell_i$s to zero.

On the other hand, in  down regulation  of a viral gene it may be desirable
to  distribute  the synonymous codon  modifications
 over many mRNA sites in order
 to reduce the chance of  spontaneous mutations yielding the original
    wild type. This is captured by Problem~\ref{prob:sdg} when
 we set the~$\ell_i$s to small non-zero  values, as then  an optimal solution will include a  transition rate reduction in many  sites.

\subsection{A biological example}
To demonstrate how the results above can be used to analyze translation  and provide 
guidelines for re-engineering the mRNA,  
  we consider the {\em S. cerevisiae}
 gene {\em YBL025W}  that encodes  the protein {\em RRN10} which is related to regulation of RNA polymerase I. This gene has $145$ codons (excluding the stop codon). Similarly to the approach used in \cite{reuveni}, we divided this
 mRNA into  $6$ consecutive pieces: the first piece includes the first $24$ codons (that are also related to later
 stages of initiation~\cite{Tuller2015}). The other pieces  include~$25$ non-overlapping codons each, 
except for the  last one that includes~$21$ codons.

To model this using an RFM with~$n=5$ sites, we first estimated
the  elongation rates~$\lambda_1,\dots,\lambda_5$   using 
   ribo-seq data for the
 codon decoding rates~\cite{Dana2014B},
 normalized so that the median elongation  rate of all {\em S. cerevisiae} mRNAs
becomes~$6.4$ codons per second \cite{Karpinets2006}. The site rate is~$(\text{site time})^{-1}$, where  
 site time  
 is the sum over the decoding times of all the codons   in this site.
These rates thus  depend on
various factors including  availability of tRNA molecules, amino acids, Aminoacyl tRNA synthetase activity and concentration, and local mRNA folding~\cite{Dana2014B,Alberts2002,Tuller2015}. 
Note that if we replace a codon in a  site of mRNA by a synonymous slower codon  then the decoding time
increases and thus the rate associated  with  this site  decreases.

The initiation rate (that corresponds to the first piece)  was estimated based
 on the ribosome density per mRNA levels, as this value is expected to be approximately proportional to 
the initiation rate when initiation is rate limiting \cite{reuveni,HRFM_steady_state}. Again we
applied a normalization that brings the median initiation rate of all {\em S. cerevisiae} mRNAs to be $0.8$ \cite{Chu2014}.
 Adding the initiation time~($1/0.4482$) to the site time of the first piece yields an RFM model with~$n=5$   and  parameters:
\[\begin{bmatrix}\bar \lambda_0 & \dots &\bar \lambda_5 \end{bmatrix} =
\begin{bmatrix} 0.1678  &
0.2572  &  0.2758  &   0.2514   &  0.2612  &   0.3002 \end{bmatrix}.
\]
 A calculation yields that
the   steady-state
production rate in this RFM is~$R = 0.0732$.

In order to analyze the solution of Problem~\ref{prob:sd} for this RFM
 we calculated the sensitivities using~\eqref{eq:sense}. This yields:
$
			s(\bar \lambda)=\begin{bmatrix}
			0.0795  &  0.0669   &  0.0611   &  0.0578  &  0.0328   &  0.0092
			\end{bmatrix} $,
so~$\bar \lambda_0$ is a bottleneck rate.  This means that the solution for Problem~\ref{prob:sd}
is to reduce all the reduction budget~$b$ from~$\bar\lambda_0$. In biological terms, this suggests that
maximal inhibition of production  should be based on
replacing some (or all) of the first~$24$ codons with  slower  synonymous codons. For comparison with the
optimization scenarios described below, consider the total budget~$b=0.0089$. The solution for Problem~~\ref{prob:sd}
is then to reduce~$\lambda_0$ by~$b$, and this yields
\be\label{eq:prodff}
R^*=0.0725.
\ee
Reducing~$\lambda_0$ by~$b$ in the model is possible by 
 substituting codons in the first site with their slowest synonymous mutation (for example, the third codon AGA should be 
 replaced by the synonymous codon~CGG, increasing the codon decoding time from~$0.1128$ seconds to~$0.2246$ seconds).

Now suppose that we are not interested in modifying these codons   
because in this region there are various regulatory signals that we may not want to change (see, for example,~\cite{Tuller2015}).
To maximize inhibition of production rate under this constraint, we 
apply Problem~\ref{prob:sdg},   with $\ell_0 = 0$, and  $\ell_i >b$  for all~$i\not =0$. Now the optimal solution is to reduce~$b$ from~$\bar \lambda_1$. Note that $\bar \lambda_1$ has the second largest sensitivity. This 
  yields $R^*=0.0726$, and is, as expected, higher than the value in~\eqref{eq:prodff}.
Again, the biological data shows that such a reduction 
  can be  done by  synonymously replacing codons $34$ (GCT with GCA), $35$ (GTT with GTA), $36$ (CCT with CCC), $38$ (CCG with CCC), $39$ (TTC with TTT), and $49$ (GTG with GTA).  

Finally, to demonstrate mutations in multiple sites, we used the data to 
find a scenario where a set of mutations yields the same total decrease in the rates. This can be done
by
synonymously replacing codons $21$ (GTG with GTA), $29$ (GAA with GAG), $58$ (TTC with TTT), $82$ (AAG with AAA), $110$ (CTA with CTG), and $141$ (GCG with GCA),  leading to
 \[
\lambda=\begin{bmatrix} 0.1677  &  0.2557 &   0.2733 &    0.2489 &    0.2599 &   0.2991 \end{bmatrix}'.
\]
 Note that all the rates are reduced and that the total reduction is $b$. This yields $R=0.0727$, which is again higher than the value in~\eqref{eq:prodff}.

 \section*{Discussion}

 There are several
 approaches for effectively  down-regulating translation. Global down-regulation  can be achieved
by  controlling  basic  translation factors
or by using  drugs that  induce
ribosome stalling  \cite{Greenberg1986,Clemens2000,Kozak1992}.
 Here we consider  down regulation of specific genes via targeting specific codons/regions in these genes. This leads to the problem of finding
the    codon regions that have the most effect on the steady-state production rate.
We study  this  problem of optimal down regulation of mRNA translation using
a mathematical model for ribosome flow, the RFM.
All possible modifications of  the rates define a feasible set
of rates, and, under certain conditions, we give a simple algorithm for finding the optimal solution, that is, the rates  that lead to a maximal decrease in the
protein production rate. For some specific cases, we also derive    theoretical results on the optimal solution.

Our results show
that the solution must   focus on the positions along the mRNA molecule where
the transition rate has the strongest
effect on the protein production rate.
However, this position is not necessarily
the one with the minimal rate (though in many cases there are correlations between the two definitions).
  Many previous studies in the field emphasized the importance of the translation bottleneck \cite{Zhang1994,Tuller2010c,Chou2004}, however,
	this is always defined as the minimal rate.   We believe that the sensitivity
  of the coding region sites should be further studied in order to  understand better
	the evolution of transcripts and their design.

The optimization problems posed here are  flexible enough to
capture various scenarios. For example, in some cases it may be desirable to
  introduce  a \emph{minimal} number of changes in the transcript to obtain the desired decrease in the translation rate.
	Indeed,  generating mutations and using suitable
	RNAi molecules  is costly in time and money.
	Also, any change
	in the translation rates can affect various important phenomena such as co-translational folding \cite{Zhang2009,Pechmann2013,Tuller2015},
	 as well as other properties that are
	 encoded in the coding region \cite{Tuller2015,Cartegni2002,Stergachis2013}.
	In other cases, such as generating a down-regulated virus strain, it may be desirable to introduce as many mutations as possible.

There are various approaches
for synthesizing  molecules that block mRNA translation
 	 (see e.g. \url{http://www.gene-tools.com/choosing_the_optimal_target}).
	In practice,  when
	determining  an optimal position to target (e.g. with RNAi molecules)
	  one must  take into account
	additional biophysical aspects. For example,
	the~GC content at the different regions along the mRNA, the folding of the mRNA, the potential binding affinity of the RNAi and the mRNA, potential un-desired binding of the RNAi to additional mRNAs or regions within the mRNA, etc. Nevertheless, we feel that out results can be
	integrated to improve  the design of such tools.

In practice, there are many mRNA molecules in the cell and they all
 compete  for the finite pool of  free   ribosomes. In particular, if more ribosomes
are stuck in a 	 traffic jam  on a certain mRNA molecule then the pool
of free ribosomes is depleted yielding a reduction in the production rates in other
mRNA molecules.
The RFM is a model for ribosome flow along a single isolated mRNA molecule.
This is a reasonable model   when the expression levels (e.g. the mRNA levels and the total number of ribosomes on  the mRNA molecules related to the gene) are relatively low, so  that changes in the translation dynamics on one mRNA have a negligible effect on the pool of ribosomes and thus on the other mRNAs.
A model for a network of~RFMs, interconnected
via a dynamic pool of free ribosomes, has been studied in~\cite{RFM_model_compete_J}.
It may be of interest to study the problem of down regulation of a specific
mRNA molecule within this framework. In this case, one can also down regulate
the mRNA indirectly by affecting the ribosomal pool. However, the tools used here do not directly apply, as the convexity results for a single chain do not
necessarily carry over to the case of a network of RFMs.

The results here suggest several biological experiments
for studying the problem of optimal down regulation and,
in particular, validating  the theoretical predictions derived using the~RFM.
Libraries encoding the same protein
using mRNAs
 with different codons (but similar mRNA levels and translation initiation rates)
can be generated as was done in \cite{Ben-Yehezkel2015}.
For each variant the  protein levels, that  are
 expected to monotonically increase with the  production rate~\cite{reuveni}, can be measured either
via a reporter protein \cite{Ben-Yehezkel2015} or directly \cite{Schwan2011}. The codon decoding rates can be estimated based on ribo-seq experiments \cite{Ben-Yehezkel2015,Dana2014B}.
Such an  experimental testbed can be used to validate
  the results reported in this study.

\section*{Appendix: Proofs}
{\sl Proof of Proposition~\ref{prop:sense}.}
Consider  Problem~\ref{prob:sd}, and suppose that~\eqref{eq:sencom}
holds. We need to show that~$\lambda^*_i=\bar \lambda_i$. Seeking a contradiction, assume that~$\lambda^*_i<\bar \lambda_i$.
By Prop.~\ref{prop:main_sol},~$\lambda^*=\bar \lambda -bd^i$, so in particular
$
R(\bar \lambda -bd^i)\leq R( \bar \lambda -bd^j).
$
Since~$R$ is a homogeneous function of the rates, we conclude that
$
R(c \bar \lambda - c bd^i)\leq R( c \bar \lambda - c bd^j)
$
for any~$c>0$.
Now taking~$c>0$ sufficiency small yields~$  \frac{\partial R(\bar \lambda)}{\partial \lambda_i}   \geq\frac{\partial R(\bar \lambda)}{\partial \lambda_j}$. This contradicts~\eqref{eq:sencom}.~\IEEEQED

{\sl Proof of Proposition~\ref{prop:cases}.	}
In the case where all the rates are equal there exists a closed-form expression for the sensitivities~\cite{RFM_sense}, namely,
\[
s_i=\frac{\sin\left(\frac{i+1}{n+3}\pi\right) \sin\left(\frac{i+2}{n+3}\pi\right)}{2(n+3)\cos^3\left(\frac{\pi}{n+3}\right)}, \quad i=0,\dots,n.
\]
This means that
\be\label{eq:cosk}
s_i= \frac{a -\cos \left(\frac{2i+3 }{n+3}\pi \right )}{b},
\ee
where~$a,b>0$ are constants that do  not depend on~$i$.
If~$n$ is even then the cosine   function in~\eqref{eq:cosk}
admits a unique minimum at~$i=n/2$, and
combining this with Proposition~\ref{prop:sense} completes the proof.
If~$n$ is odd then the cosine function in~\eqref{eq:cosk}
admits  two minima: at~$\lfloor n/2 \rfloor$ and at~$\lfloor n/2 \rfloor+1$.
Now arguing as in the proof of  Proposition~\ref{prop:sense}  and using the particle-hole symmetry of the RFM   completes the proof.~\IEEEQED

{\sl Proof of Proposition~\ref{prop:e_uniform}.	}
If $\bar e_1=\cdots=\bar e_n:=e_c$, then~\eqref{eq:ep}   yields
\be\label{eq:l_vals_un}
\bar \lambda_i=\begin{cases}
1, & i=0,\\
e_c^{-1}, & i=1,\dots,n-1,\\
e_c^{-1}-1, & i=n,
\end{cases}
\ee
where we scaled~$\bar \lambda_0$ to one w.l.o.g.
In this case, the Perron eigenvector $v\in\R^{n+2}_{++}$ of the matrix $A(\bar \lambda)$  is given by (see also~\cite{RFM_sense}):
\be\label{eq:vi_uni_e}
v_i=\begin{cases}
1, & i=1, \\
\mu^{(i-1)/2} e_c^{-1/2}, & 2 \le i \le n+1, \\
\mu^{n/2}, & i=n+2,
\end{cases}
\ee
where $\mu:=e_c/(1-e_c)$.
We consider two cases.

If~$e_c=1/2$ then $v'v=2(n+1)$ and applying Theorem~\ref{thm:spect}
yields the sensitivities:
\be\label{eq:si_un1}
s_i= \begin{cases}
\frac{1}{2(n+1)}, & i=0, \\
\frac{1}{4(n+1)}, & 1\le i \le n-1,\\
\frac{1}{2(n+1)}, & i=n.
\end{cases}
\ee
Thus, $s_0=s_n>s_j$, for all~$j\not \in\{0,n\}$, and arguing as in the proof
of Proposition~\ref{prop:sense}  and using the particle-hole symmetry implies that
the two  optimal solutions are~$\bar \lambda-bd^0$ and~$\bar \lambda-bd^n$.

If $e_c\ne 1/2$ then Theorem~\ref{thm:spect}
yields
\be\label{eq:si_un2}
s_i= \begin{cases}
\frac{1-2e_c}{1-\mu^{n+1}},  & i=0, \\
\frac{e_c(1-2e_c)}{1-\mu^{n+1}}\mu^i, & 1\le i \le n-1,\\
\frac{\mu^{n+1}(1-2e_c)}{1-\mu^{n+1}}   & i=n.
\end{cases}
\ee
When  $e_c<1/2$ [$e_c>1/2$]~\eqref{eq:si_un2} yields $s_0>s_j$, for all~$j\ne 0$ [$s_n>s_j$, for all~$j\ne n$]. Combining this with Proposition~\ref{prop:sense} completes
the proof.~\IEEEQED


\begin{thebibliography}{10}
\providecommand{\url}[1]{#1}
\csname url@rmstyle\endcsname
\providecommand{\newblock}{\relax}
\providecommand{\bibinfo}[2]{#2}
\providecommand\BIBentrySTDinterwordspacing{\spaceskip=0pt\relax}
\providecommand\BIBentryALTinterwordstretchfactor{4}
\providecommand\BIBentryALTinterwordspacing{\spaceskip=\fontdimen2\font plus
\BIBentryALTinterwordstretchfactor\fontdimen3\font minus
  \fontdimen4\font\relax}
\providecommand\BIBforeignlanguage[2]{{%
\expandafter\ifx\csname l@#1\endcsname\relax
\typeout{** WARNING: IEEEtran.bst: No hyphenation pattern has been}%
\typeout{** loaded for the language `#1'. Using the pattern for}%
\typeout{** the default language instead.}%
\else
\language=\csname l@#1\endcsname
\fi
#2}}

\bibitem{Alberts2002}
B.~Alberts, A.~Johnson, J.~Lewis, M.~Raff, K.~Roberts, and P.~Walter,
  \emph{Molecular Biology of the Cell}.\hskip 1em plus 0.5em minus 0.4em\relax
  New York: Garland Science, 2002.

\bibitem{Arava2003}
Y.~Arava, Y.~Wang, J.~D. Storey, C.~L. Liu, P.~O. Brown, and D.~Herschlag,
  ``Genome-wide analysis of {mRNA} translation profiles in {S}accharomyces
  cerevisiae,'' \emph{{Proceedings of the National Academy of Sciences}}, vol.
  100, no.~7, pp. 3889--3894, 2003.

\bibitem{Ingolia2009}
N.~T. Ingolia, S.~Ghaemmaghami, J.~R. Newman, and J.~S. Weissman,
  ``{Genome-wide analysis in vivo of translation with nucleotide resolution
  using ribosome profiling},'' \emph{Science}, vol. 324, no. 5924, pp. 218--23,
  2009.

\bibitem{Yan2016}
X.~Yan, T.~A. Hoek, R.~D. Vale, and M.~E. Tanenbaum, ``Dynamics of translation
  of single {mRNA} molecules in vivo,'' \emph{Cell}, vol. 165, no.~4, pp.
  976--89, 2016.

\bibitem{Wu2016}
B.~Wu, C.~Eliscovich, Y.~Yoon, and R.~Singer, ``Translation dynamics of single
  {mRNAs} in live cells and neurons,'' \emph{Science}, vol. 352, no. 6292, pp.
  1430--5, 2016.

\bibitem{Morisaki2016}
T.~Morisaki, K.~Lyon, K.~F. DeLuca, J.~G. DeLuca, B.~P. English, Z.~Zhang,
  L.~D. Lavis, J.~B. Grimm, S.~Viswanathan, L.~L. Looger, T.~Lionnet, and T.~J.
  Stasevich, ``Real-time quantification of single {RNA} translation dynamics in
  living cells,'' \emph{Science}, vol. 352, no. 6292, pp. 1425--9, 2016.

\bibitem{ChongWang2016}
C.~Wang, B.~Han, R.~Zhou, and X.~Zhuang, ``Real-time imaging of translation on
  single {mRNA} transcripts in live cells,'' \emph{Cell}, vol. 165, no.~4, pp.
  990--1001, 2016.

\bibitem{Ghildiyal2009}
M.~Ghildiyal and P.~Zamore, ``Small silencing {RNAs}: an expanding universe,''
  \emph{Nature Rev. Genet.}, vol.~10, pp. 94--108, 2009.

\bibitem{Inui2010}
M.~Inui, G.~Martello, and S.~Piccolo, ``{MicroRNA} control of signal
  transduction,'' \emph{Nat Rev Mol Cell Biol.}, vol.~11, no.~4, pp. 252--263,
  2010.

\bibitem{Fabian2010}
M.~Fabian, N.~Sonenberg, and W.~Filipowicz, ``Regulation of {mRNA} translation
  and stability by {microRNAs},'' \emph{Annu Rev Biochem.}, vol.~79, pp.
  351--79, 2010.

\bibitem{Filipowicz2008}
W.~Filipowicz, S.~Bhattacharyya, and N.~Sonenberg, ``Mechanisms of
  post-transcriptional regulation by {microRNAs}: are the answers in sight?''
  \emph{Nat Rev Genet.}, vol.~9, no.~2, pp. 102--14, 2008.

\bibitem{Croce2009}
C.~Croce, ``Causes and consequences of {microRNA} dysregulation in cancer,''
  \emph{Nat Rev Genet.}, vol.~10, no.~10, pp. 704--14, 2009.

\bibitem{Tavazoie2008}
S.~Tavazoie, C.~Alarc{\'o}n, T.~Oskarsson, D.~Padua, Q.~Wang, P.~Bos, and
  W.~G.~J. Massagu{\'e}, ``Endogenous human {microRNAs} that suppress breast
  cancer metastasis,'' \emph{Nature}, vol. 451, no. 7175, pp. 147--52, 2008.

\bibitem{Zhang2003}
L.~Zhang, N.~Yang, A.~Mohamed-Hadley, S.~Rubin, and G.~Coukos, ``Vector-based
  {RNAi}, a novel tool for isoform-specific knock-down of {VEGF} and
  anti-angiogenesis gene therapy of cancer,'' \emph{Biochem Biophys Res
  Commun.}, vol. 303, no.~4, pp. 1169--78, 2003.

\bibitem{Devi2006}
G.~Devi, ``{siRNA}-based approaches in cancer therapy,'' \emph{Cancer Gene
  Ther.}, vol.~13, no.~9, pp. 819--29, 2006.

\bibitem{Ben-Yehezkel2015}
T.~Ben-Yehezkel, S.~Atar, H.~Zur, A.~Diament, E.~Goz, T.~Marx, R.~Cohen,
  A.~Dana, A.~Feldman, E.~Shapiro, and T.~Tuller, ``Rationally designed,
  heterologous {S.} cerevisiae transcripts expose novel expression
  determinants,'' \emph{RNA Biol.}, vol.~12, pp. 972--84, 2015.

\bibitem{Goz2015}
E.~Goz and T.~Tuller, ``Widespread signatures of local {mRNA} folding structure
  selection in four {Dengue} virus serotypes,'' \emph{BMC Genomics}, vol.~16,
  no. Suppl 10:S4., 2015.

\bibitem{Kaspar2005}
Q.~Wang, C.~Contag, H.~Ilves, B.~Johnston, and R.~Kaspar, ``Small hairpin
  {RNAs} efficiently inhibit hepatitis {C IRES}-mediated gene expression in
  human tissue culture cells and a mouse model,'' \emph{Molecular Therapy},
  vol.~12, no.~3, pp. 562--8, 2005.

\bibitem{Coleman2008}
J.~Coleman, D.~Papamichail, S.~Skiena, B.~Futcher, E.~Wimmer, and S.~Mueller,
  ``{Virus attenuation by genome-scale changes in codon pair bias},''
  \emph{Science}, vol. 320, pp. 1784--7, 2008.

\bibitem{Perez2009}
J.~Perez, A.~Pham, M.~Lorini, M.~Chua, J.~Steel, and B.~tenOever,
  ``{MicroRNA}-mediated species-specific attenuation of influenza {A} virus,''
  \emph{Nat Biotechnol.}, vol.~27, no.~6, pp. 572--6, 2009.

\bibitem{Tuller2010c}
T.~Tuller, A.~Carmi, K.~Vestsigian, S.~Navon, Y.~Dorfan, J.~Zaborske, T.~Pan,
  O.~Dahan, I.~Furman, and Y.~Pilpel, ``{An evolutionarily conserved mechanism
  for controlling the efficiency of protein translation},'' \emph{Cell}, vol.
  141, no.~2, pp. 344--54, 2010.

\bibitem{Forman2008}
J.~Forman, A.~Legesse-Miller, and H.~Coller, ``A search for conserved sequences
  in coding regions reveals that the let-7 {microRNA} targets {Dicer} within
  its coding sequence,'' \emph{{Proceedings of the National Academy of
  Sciences}}, vol. 105, no.~39, pp. 14\,879--84, 2008.

\bibitem{Fang2004}
P.~Fang, C.~Spevak, C.~Wu, and M.~Sachs, ``A nascent polypeptide domain that
  can regulate translation elongation,'' \emph{{Proceedings of the National
  Academy of Sciences}}, vol. 101, no.~12, pp. 4059--64, 2004.

\bibitem{Shaw2003}
L.~B. Shaw, R.~K.~P. Zia, and K.~H. Lee, ``{Totally asymmetric exclusion
  process with extended objects: a model for protein synthesis},'' \emph{Phys.
  Rev. E}, vol.~68, p. 021910, 2003.

\bibitem{TASEP_tutorial_2011}
R.~Zia, J.~Dong, and B.~Schmittmann, ``{Modeling translation in protein
  synthesis with \mbox{TASEP}: A tutorial and recent developments},'' \emph{J.
  Statistical Physics}, vol. 144, pp. 405--428, 2011.

\bibitem{TASEP_book}
A.~Schadschneider, D.~Chowdhury, and K.~Nishinari, \emph{{Stochastic Transport
  in Complex Systems: From Molecules to Vehicles}}.\hskip 1em plus 0.5em minus
  0.4em\relax Elsevier, 2011.

\bibitem{tasep_ad_hoc_nets}
S.~Srinivasa and M.~Haenggi, ``{A statistical mechanics-based framework to
  analyze ad hoc networks with random access},'' \emph{IEEE Trans. Mobile
  Computing}, vol.~11, pp. 618--630, 2012.

\bibitem{reuveni}
S.~Reuveni, I.~Meilijson, M.~Kupiec, E.~Ruppin, and T.~Tuller, ``{Genome-scale
  analysis of translation elongation with a ribosome flow model},'' \emph{{PLOS
  Computational Biology}}, vol.~7, p. e1002127, 2011.

\bibitem{solvers_guide}
R.~A. Blythe and M.~R. Evans, ``Nonequilibrium steady states of matrix-product
  form: a solver's guide,'' \emph{J. Phys. A: Math. Theor.}, vol.~40, no.~46,
  pp. R333--R441, 2007.

\bibitem{RFM_stability}
M.~Margaliot and T.~Tuller, ``{Stability analysis of the ribosome flow
  model},'' \emph{{IEEE/ACM Trans. Computational Biology and Bioinformatics}},
  vol.~9, pp. 1545--1552, 2012.

\bibitem{Diament2016}
A.~Diament and T.~Tuller, ``Ribosome profiling resolution in practice,''
  \emph{under review}, 2016.

\bibitem{Kaern2005}
M.~Kaern, T.~C. Elston, W.~J. Blake, and J.~J. Collins, ``Stochasticity in gene
  expression: from theories to phenotypes,'' \emph{Nat Rev Genet.}, vol.~6, pp.
  451--64, 2005.

\bibitem{hlsmith}
H.~L. Smith, \emph{{Monotone Dynamical Systems: An Introduction to the Theory
  of Competitive and Cooperative Systems}}, ser. {Mathematical Surveys and
  Monographs}.\hskip 1em plus 0.5em minus 0.4em\relax Providence, RI: Amer.
  Math. Soc., 1995, vol.~41.

\bibitem{RFM_entrain}
M.~Margaliot, E.~D. Sontag, and T.~Tuller, ``{Entrainment to periodic
  initiation and transition rates in a computational model for gene
  translation},'' \emph{PLoS ONE}, vol.~9, no.~5, p. e96039, 2014.

\bibitem{RFM_max}
G.~Poker, Y.~Zarai, M.~Margaliot, and T.~Tuller, ``Maximizing protein
  translation rate in the nonhomogeneous ribosome flow model: A convex
  optimization approach,'' \emph{J. Royal Society Interface}, vol.~11, no. 100,
  p. 20140713, 2014.

\bibitem{zarai_infi}
Y.~Zarai, M.~Margaliot, and T.~Tuller, ``Explicit expression for the
  steady-state translation rate in the infinite-dimensional homogeneous
  ribosome flow model,'' \emph{{IEEE/ACM Trans. Computational Biology and
  Bioinformatics}}, vol.~10, pp. 1322--1328, 2013.

\bibitem{RFM_sense}
G.~Poker, M.~Margaliot, and T.~Tuller, ``Sensitivity of {mRNA} translation,''
  \emph{Sci. Rep.}, vol.~5, no. 12795, 2015.

\bibitem{RFM_feedback}
{Margaliot, M. and Tuller, T.}, ``{Ribosome flow model with positive
  feedback},'' \emph{J. Royal Society Interface}, vol.~10, p. 20130267, 2013.

\bibitem{RFMR}
A.~Raveh, Y.~Zarai, M.~Margaliot, and T.~Tuller, ``Ribosome flow model on a
  ring,'' \emph{{IEEE/ACM Trans. Computational Biology and Bioinformatics}},
  vol.~12, no.~6, pp. 1429--1439, 2015.

\bibitem{rfm_control}
\BIBentryALTinterwordspacing
Y.~Zarai, M.~Margaliot, E.~D. Sontag, and T.~Tuller, ``Controlling {mRNA}
  translation,'' 2016, {Submitted}. [Online]. Available:
  \url{http://arxiv.org/abs/1602.02308}
\BIBentrySTDinterwordspacing

\bibitem{RFM_model_compete_J}
A.~Raveh, M.~Margaliot, E.~D. Sontag, and T.~Tuller, ``A model for competition
  for ribosomes in the cell,'' \emph{J. Royal Society Interface}, vol.~13, no.
  116, 2016.

\bibitem{Dana2014B}
A.~Dana and T.~Tuller, ``Mean of the typical decoding rates: a new translation
  efficiency index based on ribosome analysis data,'' \emph{G3: Genes, Genomes,
  Genetics}, 2014.

\bibitem{TullerGB2011}
T.~Tuller, I.~Veksler, N.~Gazit, M.~Kupiec, E.~Ruppin, and M.~Ziv, ``Composite
  effects of the coding sequences determinants on the speed and density of
  ribosomes,'' \emph{Genome Biol.}, vol.~12, no.~11, p. R110, 2011.

\bibitem{Sabi2015}
R.~Sabi and T.~Tuller, ``A comparative genomics study on the effect of
  individual amino acids on ribosome stalling,'' \emph{BMC Genomics}, vol.~16,
  p.~S5, 2015.

\bibitem{beck2014}
A.~Beck, \emph{Introduction to Nonlinear Optimization: Theory, Algorithms, and
  Applications with MATLAB}.\hskip 1em plus 0.5em minus 0.4em\relax
  Philadelphia, PA: Society for Industrial and Applied Mathematics, 2014.

\bibitem{concave_prog}
R.~Enhbat, ``An algorithm for maximizing a convex function over a simple set,''
  \emph{J. Global Optimization}, vol.~8, no.~4, pp. 379--391, 1996.

\bibitem{Zhang2009}
G.~Zhang, M.~Hubalewska, and Z.~Ignatova, ``Transient ribosomal attenuation
  coordinates protein synthesis and co-translational folding,'' \emph{Nat
  Struct Mol Biol.}, vol.~16, no.~3, pp. 274--80, 2009.

\bibitem{Pechmann2013}
S.~Pechmann and J.~Frydman, ``Evolutionary conservation of codon optimality
  reveals hidden signatures of cotranslational folding,'' \emph{Nat Struct Mol
  Biol.}, vol.~20, no.~2, pp. 237--43, 2013.

\bibitem{Tuller2015}
T.~Tuller and H.~Zur, ``Multiple roles of the coding sequence 5' end in gene
  expression regulation,'' \emph{Nucleic Acids Res.}, vol.~43, no.~1, pp.
  13--28, 2015.

\bibitem{Zhang2000}
J.~Zhang, ``Protein-length distributions for the three domains of life,''
  \emph{Trends Genet.}, vol.~16, no.~3, pp. 107--9, 2000.

\bibitem{phos_relays}
A.~Csikasz-Nagy, L.~Cardelli, and O.~S. Soyer, ``Response dynamics of
  phosphorelays suggest their potential utility in cell signaling,'' \emph{J.
  Royal Society Interface}, vol.~8, pp. 480--488, 2011.

\bibitem{toward_prod_rates}
J.~J. Dong, B.~Schmittmann, and R.~K.~P. Zia, ``Towards a model for protein
  production rates,'' \emph{J. Statistical Physics}, vol. 128, no. 1-2, pp.
  21--34, 2007.

\bibitem{edge_tasep_2009}
J.~J. Dong, R.~K.~P. Zia, and B.~Schmittmann, ``Understanding the edge effect
  in {TASEP} with mean-field theoretic approaches,'' \emph{J. Phys. A: Math.
  Gen.}, vol.~42, no.~1, p. 015002, 2009.

\bibitem{PhysRevE.76.051113}
J.~J. Dong, B.~Schmittmann, and R.~K.~P. Zia, ``Inhomogeneous exclusion
  processes with extended objects: The effect of defect locations,''
  \emph{Phys. Rev. E}, vol.~76, p. 051113, 2007.

\bibitem{foulaadvand2008asymmetric}
M.~E. Foulaadvand, A.~B. Kolomeisky, and H.~Teymouri, ``Asymmetric exclusion
  processes with disorder: Effect of correlations,'' \emph{Physical Review E},
  vol.~78, no.~6, p. 061116, 2008.

\bibitem{Chou2004}
T.~Chou and G.~Lakatos, ``Clustered bottlenecks in {mRNA} translation and
  protein synthesis,'' \emph{Phys. Rev. Lett.}, vol.~93, p. 198101, 2004.

\bibitem{PhysRevE.58.1911}
G.~Tripathy and M.~Barma, ``Driven lattice gases with quenched disorder: exact
  results and different macroscopic regimes,'' \emph{Phys. Rev. E}, vol.~58,
  pp. 1911--1926, 1998.

\bibitem{Karpinets2006}
T.~V. Karpinets, D.~J. Greenwood, C.~E. Sams, and J.~T. Ammons, ``{RNA}:protein
  ratio of the unicellular organism as a characteristic of phosphorous and
  nitrogen stoichiometry and of the cellular requirement of ribosomes for
  protein synthesis,'' \emph{BMC Biol.}, vol.~4, no.~30, pp. 274--80, 2006.

\bibitem{HRFM_steady_state}
M.~Margaliot and T.~Tuller, ``{On the steady-state distribution in the
  homogeneous ribosome flow model},'' \emph{{IEEE/ACM Trans. Computational
  Biology and Bioinformatics}}, vol.~9, pp. 1724--1736, 2012.

\bibitem{Chu2014}
D.~Chu, E.~Kazana, N.~Bellanger, T.~Singh, M.~F. Tuite, and T.~von~der Haar,
  ``Translation elongation can control translation initiation on eukaryotic
  {mRNAs},'' \emph{EMBO J.}, vol.~33, no.~1, pp. 21--34, 2014.

\bibitem{Greenberg1986}
M.~Greenberg, A.~Hermanowski, and E.~Ziff, ``Effect of protein synthesis
  inhibitors on growth factor activation of c-fos, c-myc, and actin gene
  transcription,'' \emph{Mol Cell Biol.}, vol.~6, no.~4, pp. 1050--7, 1986.

\bibitem{Clemens2000}
M.~Clemens, M.~Bushell, I.~Jeffrey, V.~Pain, and S.~Morley, ``Translation
  initiation factor modifications and the regulation of protein synthesis in
  apoptotic cells,'' \emph{Cell Death Differ.}, vol.~7, no.~7, pp. 603--15,
  2000.

\bibitem{Kozak1992}
M.~Kozak, ``Regulation of translation in eukaryotic systems,'' \emph{Annu Rev
  Cell Biol.}, vol.~8, pp. 197--225, 1992.

\bibitem{Zhang1994}
S.~Z. Zubay, E.~Goldman, and G., ``{Clustering of low usage codons and ribosome
  movement},'' \emph{J. Theor. Biol.}, vol. 170, pp. 339--54, 1994.

\bibitem{Cartegni2002}
L.~Cartegni, S.~Chew, and A.~Krainer, ``Listening to silence and understanding
  nonsense: exonic mutations that affect splicing,'' \emph{Nat. Rev. Genet.},
  vol.~3, pp. 285--98, 2002.

\bibitem{Stergachis2013}
A.~Stergachis, E.~Haugen, A.~Shafer, W.~Fu, B.~Vernot, A.~Reynolds,
  A.~Raubitschek, S.~Ziegler, E.~LeProust, J.~Akey, and J.~Stamatoyannopoulos,
  ``Exonic transcription factor binding directs codon choice and affects
  protein evolution,'' \emph{Science}, vol. 342, pp. 1367--72, 2013.

\bibitem{Schwan2011}
B.~Schwanhausser, D.~Busse, N.~Li, G.~Dittmar, J.~Schuchhardt, J.~Wolf,
  W.~Chen, and M.~Selbach, ``Global quantification of mammalian gene expression
  control,'' \emph{Nature}, vol. 473, no. 7347, pp. 1367--72, 2011.

\end{thebibliography}

     \end{document}